\documentclass[aps,pra,12pt,groupedaddress,showpacs]{revtex4}
\usepackage{amssymb}
\usepackage{amsfonts}
\usepackage{amsmath}
\usepackage{graphicx}

\newtheorem{example}{Example}[section]

\newtheorem{theorem}[example]{Theorem}

\def\<{\langle}
\def\>{\rangle}

\def\N{{\mathbb N}}
\def\C{{\mathbb C\, }}

\def\Q{{\mathbb Q}}

\def\PP{{\mathbb P}}
\def\xx{{\bf x}}
\def\yy{{\bf y}}
\def\zz{{\bf z}}

\def\bbf#1{{\bf #1}}

\begin{document}

\title{The moduli space of three qutrit states}

\author{Emmanuel Briand}
\email{ebriand@univ-mlv.fr}
\author{Jean-Gabriel Luque}
\email{luque@univ-mlv.fr}
\author{Jean-Yves Thibon}
\email{jyt@univ-mlv.fr}
\affiliation{Institut Gaspard Monge, Universit\'e de Marne-la-Vall\'ee\\
F-77454 Marne-la-Vall\'ee cedex, France}
\author{Frank Verstraete}
\email{frank.verstraete@mpq.mpg.de}
\affiliation{Max-Planck-Institut f\"ur QuantenOptik,
Hans-Kopfermann-Str.~1\\D-85748 Garching, Germany}

\begin{abstract}
We study the invariant theory of trilinear forms over a three-dimensional
complex vector space, and apply it to investigate  
the behaviour of pure entangled three-partite qutrit states and
their normal forms under
local filtering operations (SLOCC).
We describe the orbit space of the SLOCC group
$SL(3,\C)^{\times 3}$ both in its affine and
projective versions in terms of a very symmetric normal form
parameterized by three complex
numbers. The parameters of the possible normal forms of a given state
are roots of an algebraic equation, which
is proved to be solvable by radicals. The structure of the
sets of equivalent normal forms is related to the geometry
of certain regular complex polytopes.
\end{abstract}


\maketitle
\newpage
\section{\label{intro}Introduction}
The invariant theory of trilinear forms over a three-dimensional complex vector
space is an old subject with a long history, which, as we shall see, appears even
longer if we take into account certain indirect but highly relevant contributions
\cite{Ma,OS,ShTo,Cox}. This question has been recently revived in the field of
Quantum Information Theory as the problem of classifying entanglement patterns of
three-qutrit states.

Indeed, since the advent of quantum computation and quantum cryptography, entanglement has
been promoted to a resource that allows quantum physics to perform tasks that are
classically impossible. Quantum cryptography \cite{BB84,Eke91} proved that this
gap even exists with small systems of two entangled qubits. Furthermore, it is
expected that the study of higher dimensional systems and of multipartite (e.g.
3-partite) states would lead to more applications. A seminal example is the
so-called 3-qutrit Aharonov-state, which ``\emph{is so elegant it had to be
useful}"\cite{FGM01}: Fitzi, Gisin and Maurer \cite{FGM01} found out that the
classically impossible Byzantine agreement problem \cite{Lamportseegisin}  can be
solved using 3-partite qutrit states. From a more fundamental point a view, the
Aharonov state led to non-trivial counterexamples of the conjectures on additivity
of the relative entropy of entanglement \cite{VW01} and of the output purity of
quantum channels \cite{WH02}. Obviously, these results provide a strong motivation
for studying 3-partite qutrit states. Furthermore, interesting families of
higher-dimensional states are perfectly suited to address questions concerning
local realism and Bell inequalities (see e.g. \cite{KGL02} for a study of
three-qutrit correlations).

It is therefore of interest to find some classification scheme for three-qutrit
states. A possible direction is to look for classes of equivalent states, in the
sense that they are equivalent up to local unitary transformations
\cite{AAC00,CHS00,VDD01a} or local filtering operations (also called SLOCC
operations) \cite{DVC00,VDD01a,Klyachko,Ver,LT1}. In the case of three qubits,
especially the last classification proved to yield a lot of insights (the
classification up to local unitaries has too much parameters left); the reason for
that is that in the closure of each generic orbit induced by SLOCC operations,
there is a unique state (up to local unitary transformations) with maximal
entanglement \cite{VDD01a,Klyachko}.

In \cite{Ver}, a numerical method converging to such a maximally entangled state
has been described. It has been experimentally observed that, when applied to a
three qutrit state, this method converged to a very special normal form. We shall
provide a formal proof of this property, and then study in some detail the
geometry of those normal forms. Precise statements of the results are summarized
in the forthcoming section.

\section{\label{results}Results}
Let $V=\C^3$  and ${\cal H}=V\otimes V\otimes V$ regarded as a
representation of the group $G=SL(3,\C)^{\times 3}$.
 The elements
of ${\cal H}$ will be interpreted either as three-qutrit states

\begin{equation}\label{3qutrit}
|\psi\rangle=\sum_{i,j,k=1}^3A_{ijk}|i,j,k\rangle
\end{equation}

\noindent or as a trilinear forms

\begin{equation}
f=f(\bbf x,\bbf y,\bbf z)=\sum_{i,j,k=1}^3A_{ijk}x_iy_jz_k
\end{equation}

\noindent that is, we identify the basis state $|ijk\rangle$ with the
monomial $x_iy_jz_k$. If $g=(g^{(1)},g^{(2)},g^{(3)})\in G$ is a
triple of matrices, we define $x'_i=\sum_pg_{ip}^{(1)}x_p$,
$y'_j=\sum_qg_{jq}^{(2)}y_q$,$z'_k=\sum_rg_{kr}^{(3)}x_r$, and the
coefficients $A'_{ijk}$ by the condition

\begin{equation}
\sum A_{ijk}'x'_iy'_jz'_k=\sum A_{ijk}x_iy_jz_k,
\end{equation}
the action of $G$ on $\cal H$ being defined by
\begin{equation}
g\cdot f=\sum A'_{ijk}x_iy_jz_k
\end{equation}
It has been shown by Vinberg \cite{Vin1} that a generic state can
be reduced to the normal form

\begin{equation}\label{normalform}
A'_{ijk}=u\delta_{ijk}+{w-v\over 2}\epsilon_{ijk}+{w+v\over
2}|\epsilon_{ijk}|
\end{equation}
(where $\delta_{ijk}$ is the Kronecker symbol and $\epsilon_{ijk}$ the
completely antisymetric tensor)
by an appropriate choice of $g\in G$.

 Our first result is

 \begin{theorem}\label{conv}
When applied to a generic $3$-qutrit state (\ref{3qutrit}) the numerical algorithm of
\cite{Ver} converges to a state which is a Vinberg normal
form,   generically in the same $G$-orbit as $|\psi\rangle$.
 \end{theorem}

 As proved in \cite{Ver,VDD01a}, the normal form $|\psi'\rangle$ is
 unique up to local unitary transformations.
 More precisely,

 \begin{theorem}\label{thnormalform}
A generic state has exactly $648$ different normal forms. For
special states, this number can be reduced to $216$, $72$, $27$ or
$1$. Moreover, the coefficients $u$, $v$, $w$ of the normal form
can be computed algebraically.
 \end{theorem}

 \begin{theorem}\label{coef}
The coefficients of the normal forms are determined, up to a sign, by an algebraic
equation of degree $1296$, which is explicitly solvable by
radicals.
 \end{theorem}
To form this equation, we need some notions of
invariant theory.

A polynomial $P(A)$ in the coefficients $A_{ijk}$ is an
{\em invariant} of the action of $G$ on $\cal H$ if $P(A')=P(A)$ for all $g\in G$.
These invariants form a graded algebra $R$ (any invariant $P$ is a
sum of homogeneous invariants) and the first issue is to
determine the dimension of the space $R_d$ of homogeneous
invariant of degree $d$.
The Hilbert series
\begin{equation}\label{hilbert1}
h(t)=\sum_{d\geq 0}\dim R_dt^d
\end{equation}
is known \cite{Vin1}

\begin{equation}\label{hilb}
h(t)={1\over(1-t^6)(1-t^9)(1-t^{12})}
\end{equation}
and in fact, one can prove that $R$ is a polynomial algebra
generated by three algebraically independent invariants of
respective degree $6$, $9$ and $12$.

The modern way to prove this result is due to Vinberg, who
obtained it from his notion of Weyl group of a graded Lie algebra,
applied to a $Z_3$-grading of the exceptional Lie algebra $E_6$
 \cite{Vin2}.

In section \ref{inv}, we shall explain how it can be deduced from
the work of Chanler \cite{Chan}. We prove that certain invariants $I_6$, $I_9$
and $I_{12}$ introduced  in Ref. \cite{Chan} are indeed
algebraic generators of $R$  and explain how to compute them from
the numerical values of the coefficients $A_{ijk}$, by expressing them
in terms of {\em transvectants}, that is, by means of certain differential
polynomials in the form $f$, rather than in terms of the classical
symbolic notation. Given the values of the invariants for a
particular state, we show how to form and solve the system of
algebraic equations determining the coefficients, $u,v, w$ of the
normal form.

Let $a=I_6$, $b=I_{12}$ and $c=I_{18}$ (a certain polynomial
in the fundamental invariants).
Then, the symmetric functions of  $u^3$, $v^3$ and $w^3$
\begin{equation}
\psi=u^3+v^3+w^3, \chi=u^3v^3+u^3w^3+v^3w^3, \lambda=216u^3v^3w^3
\end{equation}
satisfy
\begin{equation}\label{system}
\left\{\begin{array}{l}
\psi^2-12\chi-a=0\\
\psi^4+\lambda\psi-b=0\\ \psi^6-{5\over 2}\lambda\psi^3-{1\over
8}\lambda^2-c=0\\
\end{array}\right.
\end{equation}

\begin{theorem}\label{rsystem}
The system (\ref{system}) has generically $1296$ solutions
$(u,v,w)$, which can be obtained by solving a chain of algebraic
equations of degree at most $4$. Only $648$ of them give the correct sign for
$I_9$.
The number of solutions (with the correct sign for $I_9$) can be reduced only to $216$, $72$, $27$ or
$1$. Moreover, the isotropy groups of these degenerate orbits can
be determined, and the configuration of the points $(u,v,w)$ in $\C^3$ can be
interpreted in terms of the
geometry of regular complex polyhedra.
\end{theorem}
The details are given in Section
\ref{formprobl}.

\section{\label{inv}The fundamental invariants}

In this section, we describe the fundamental
invariants, as well as the other concomitants obtained by Chanler \cite{Chan},
in a form  suitable for calculations, in particular for their
numerical evaluation.

As already mentioned, we shall identify a three qutrit state $|\psi\rangle\in{\cal H}$
with a trilinear form
\begin{equation}
f(\xx,\yy,\zz)=\sum_{1\leq i,j,k\leq 3}A_{ijk}x_iy_jz_k
\end{equation}
in three ternary variables.
To construct its fundamental invariants, we shall need
the notion of a transvectant,  which is defined by means
of Cayley's Omega process (see, e.g., Ref. \cite{Turn}).

Let $f_1$, $f_2$ and $f_3$ be three forms  in a ternary variable
${\bf x}=(x_1,x_2,x_3)$. Their tensor product $f_1\otimes
f_2\otimes f_3$ is identified  with the polynomial $f_1({\bf
x}^{(1)})f_2({\bf x}^{(2)})f_3({\bf x}^{(3)})$ in the three
independent
ternary variables ${\bf x}^{(1)}$, ${\bf x}^{(2)}$ and ${\bf
x}^{(3)}$. We use the ``trace'' notation  of Olver \cite{Olver} to denote
the multiplication map $f_1\otimes f_2\otimes f_3\rightarrow
f_1f_2f_3$, that is,
\begin{equation}
{\rm tr } f_1({\bf x}^{(1)})f_2({\bf x}^{(2)})f_3({\bf
x}^{(3)})=f_1({\bf x})f_2({\bf x})f_3({\bf x})
\end{equation}
Cayley's operator $\Omega_{\bf x}$ is the differential
operator
\begin{equation}
\Omega_{\bf x}=\left|\begin{array}{ccc}\partial\over\partial
x_1^{(1)}&\partial\over\partial x_1^{(2)}&\partial\over\partial
x_1^{(3)}\\\partial\over\partial x_2^{(1)}&\partial\over\partial
x_2^{(2)}&\partial\over\partial x_2^{(3)}\\\partial\over\partial
x_3^{(1)}&\partial\over\partial x_3^{(2)}&\partial\over\partial
x_3^{(3)}\end{array}\right|\end{equation}
Now, we consider three
independent ternary variables $\bf x$, $\bf y$ and $\bf z$
together with the associated dual (contravariant) variables
${\bf \xi}=(\xi_1,\xi_2,\xi_3)$,
${\bf \eta}=(\eta_1,\eta_2,\eta_3)$, ${\bf \zeta}=(\zeta_1,\zeta_2,\zeta_3)$
(that is, $\xi_i$ is the linear form on the ${\bf x}$ space
such that $\xi_i(x_j)=\delta_{ij}$).

A {\em concomitant} of $f$ is, by definition,  a polynomial
$F$ in the $A_{ijk}$, $\bf x$, $\bf y$, $\bf z$, ${\bf \xi}$,
${\bf \eta}$, ${\bf \zeta}$, such that if
 $g=(g_1, g_2, g_3)\in
SL(3,\C)^3$, then, with $A'$, $\bf x'$ etc. as above,
\begin{eqnarray}
F(A';{\bf x'},{\bf y'},{\bf
z'};\xi',\eta',\zeta')=
F(A;{\bf
x},{\bf y},{\bf z};\xi,\eta,\zeta).
\end{eqnarray}

The
algebra of concomitants admits only one generator of degree $1$ in
the $A_{ijk}$, which is the form $f$ itself. Other concomitants
can be deduced from $f$ and the three absolute invariants
$P_\alpha=\sum\xi_ix_i$, $P_\beta=\sum \eta_jy_j$ and
$P_\gamma=\sum\zeta_kz_k$, using transvectants. If $F_1$, $F_2$
and $F_3$ are three $6$-tuple forms in the independent ternary
variables ${\bf x}$, ${\bf y}$, ${\bf z}$, ${\bf \xi}$, $\bf \eta$
and $\bf \zeta$, one defines for any
$(n_1,n_2,n_3)\times(m_1,m_2,m_3)\in\N^3\times\N^3$ the multiple
transvectant of $F_1$, $F_2$ and $F_3$ by
 \begin{eqnarray}
(F_1,F_2,F_3)^{n_1n_2n_3}_{m_1m_2m_3}={\rm tr }\quad\Omega_{\bf
x}^{n_1}\Omega_{\bf y}^{n_2}\Omega_{\bf
z}^{n_3}\Omega_{\xi}^{m_1}\Omega_\eta^{m_2}\Omega_\zeta^{m_3}\nonumber
\\ \prod_{i=1}^3F_i({\bf
x}^{(i)},{\bf y}^{(i)},{\bf z}^{(i)};
{\bf \xi}^{(i)},\eta^{(i)},\zeta^{(i)}).\end{eqnarray}
For convenience, we will set
$(F_1,F_2,F_3)^{n_1n_2n_3}=(F_1,F_2,F_3)^{n_1n_2n_3}_{000}$. The
concomitants of degree $2$ given by Chanler \cite{Chan} can be
obtained using these operations:
\begin{eqnarray}
Q_\alpha=(f,f,P_\beta P_\gamma)^{011}\\Q_\beta=(f,f,P_\alpha
P_\gamma)^{101}\\Q_\gamma=(f,f,P_\alpha P_\beta)^{110}
\end{eqnarray}
The invariant $I_6$ is then
\begin{equation}
I_6=\frac1{96}(Q_\alpha,Q_\alpha,Q_\alpha)^{200}_{011}=
\frac1{96}(Q_\beta,Q_\beta,Q_\beta)^{020}_{101}=
\frac1{96}(Q_\gamma,Q_\gamma,Q_\gamma)^{002}_{110}.
\end{equation}
There is  an alternative expression using only the ground form $f$,
\begin{equation}
I_6=\frac1{1152}(f^2,f^2,f^2)^{222}.
\end{equation}
Now, in degree $3$ the covariants $B_\alpha$, $B_\beta$ and
$B_\gamma$ of Ref. \cite{Chan} are
\begin{eqnarray}
B_\alpha=(f,f,f)^{011}\\B_\beta=(f,f,f)^{101}\\B_\gamma=(f,f,f)^{110}.
\end{eqnarray}
The other concomitants found by Chanler can be written in a
similar way:
\begin{eqnarray}
C_{\alpha\beta}&=&\frac14(f,f,fP_\beta)^{110}\\
C_{\beta\alpha}&=&\frac14(f,f,fP_\alpha)^{110}\\
C_{\alpha\gamma}&=&\frac14(f,f,fP_\gamma)^{101}\\
C_{\gamma\alpha}&=&\frac14(f,f,fP_\alpha)^{101}\\
C_{\beta\gamma}&=&\frac14(f,f,fP_\gamma)^{011}\\
C_{\gamma\beta}&=&\frac14(f,f,fP_\beta)^{011}\\
D_\alpha&=&-2(fP_\beta,fP_\gamma,f)^{111}\\
D_\beta&=&2(fP_\alpha,fP_\gamma,f)^{111}\\
D_\gamma&=&-2(fP_\alpha,fP_{\beta},f)^{111}\\
E_\alpha&=&(Q_\alpha,f,P_\alpha)^{100}\\
E_\beta&=&(Q_\beta,f,P_\beta)^{010}\\
E_\gamma&=&(Q_\gamma,f,P_\gamma)^{001}\\
G_{\alpha}&=&-\frac38(fP_\beta,fP_\gamma,f)^{011}+\frac5{16}(fP_\beta
 P_\gamma,f,f)^{011}\\
G_{\beta}&=&-\frac3{8}(fP_\alpha,fP_\gamma,f)^{101}+\frac5{16}(fP_\alpha
 P_\gamma,f,f)^{101}\\
G_{\gamma}&=&-\frac3{8}(fP_\alpha,fP_\beta,f)^{110}+\frac5{16}(fP_\alpha
 P_\beta,f,f)^{110}\\
H&=&\frac12(fP_\alpha,fP_\beta,f P_\gamma)^{111}
\end{eqnarray}
Here,  we have combined the concomitants of degree
$0$, $1$ and $2$ into independent concomitants of
degree $3$. Next, we have chosen the scalar factors so that the
syzygies given by Chanler \cite{Chan} hold in the form
\begin{eqnarray}
H+E_\alpha-E_\gamma+D_\beta P_\beta=0,\\
 H+E_\beta-E_\alpha+D_\gamma
P_\gamma=0,\\
 H+E_\gamma-E_\beta+D_\alpha P_\alpha=0,
\\
3C_{\alpha\beta}-B_\gamma P_\beta=0,\\
 3C_{\beta\alpha}-B_\gamma P_\alpha=0, \\
3C_{\alpha\gamma}-B_\beta P_\gamma=0,\\
3C_{\gamma\alpha}-B_\beta P_\alpha=0,\\
 3C_{\beta\gamma}-B_\alpha
P_\gamma=0,\\
3C_{\gamma\beta}-B_\alpha P_\beta=0,
\\
6G_\alpha-3Q_\alpha f+B_\alpha P_\beta P_\gamma=0,\\
6G_\beta-3Q_\beta f+B_\beta P_\alpha P_\gamma=0,\\
6G_\gamma-3Q_\gamma f+B_\gamma P_\alpha P_\beta=0.
\end{eqnarray}
One can remark that a basis of the space of the
concomitants of degree $3$ found by Chanler can be constructed
using only transvections and products from smaller degrees,
\begin{equation}
f^3,\, Q_\alpha f,\, Q_\beta f, Q_\gamma f, B_\alpha, B_\beta,
B_\gamma, D_\alpha, D_\beta, D_\gamma, E_\alpha.
\end{equation}
The knowledge of these concomitants allows
one to construct the invariants $I_9$ and $I_{12}$
\begin{eqnarray}
I_9&=&\frac1{576}(E_\alpha,E_\beta,E_\beta)^{111}_{111}\\
I_{12}&=&\frac1{124416}(B_\alpha f,B_\alpha f,B_\alpha f)^{411}.
\end{eqnarray}
These expressions, which can be easily implemented in any computer algebra system,
 will prove convenient to compute the specializations
discussed in the sequel.

\section{\label{genNF}Normal form and invariants}

It will now be shown that a generic state can be reduced to the normal form
\begin{equation}
A_{ijk}= u\delta_{ijk}+ \frac{w-v}{2}\epsilon_{ijk}+\frac{w+v}{2}|\epsilon_{ijk}|
\end{equation}
where $\epsilon_{ijk}$ is the alternating tensor,
or, otherwise said, that the generic trilinear form $f(\xx,\yy,\zz)$ is equivalent
to some
\begin{eqnarray}
N_{uvw}(\xx,\yy,\zz) &= u(x_1y_1z_1+x_2y_2z_2 +x_3y_3z_3)\nonumber\\
&+ v(x_1y_3z_2+x_2y_1z_3 +x_3y_2z_1)\\
&+w(x_1y_2z_3+x_2y_3z_1 +x_3y_1z_2)\nonumber
\end{eqnarray}
For such a state, the local density operators are all proportional
to the identity. This property is automatically satistfied by
the limiting state obtained from the numerical method of
Ref. \cite{Ver}, and implies maximal entanglement as well.
Since this algorithm amounts to an infinite sequence
of invertible local filtering operations,
the genericity of Vinberg's normal form, together with the
previously mentioned properties, implies convergence to
a Vinberg normal form for a generic input state, that is, our theorem \ref{conv}.

This normal form is in general not unique, and the relations between the
various $N_{uvw}$ in a given orbit is an interesting question, which will be
addressed in the sequel.

Although  the validity of this normal form follows from Vinberg's theory \cite{Vin2},
it can also  be proved in  other ways, some of them being
particularly instructive.
We shall detail one of these possibilities, which  will give us the
opportunity to introduce some important polynomials,
playing a role in the algebraic calculation of the normal
form and in the geometric discussion of the orbits.

The shortest possibility,
although not the most elementary,
relies on the results of Ref. \cite{Chan}, and
starts with computing the
invariants of $N_{uvw}$. We then use a few results of algebraic geometry, which can
be found in \cite{Encycl}. Let us denote by $C_k\equiv C_k(u,v,w)$ ($k=6,9,12$)
the values of the $I_k$ on $N_{uvw}$. Direct calculation gives, denoting
by $m_{pqr}$
the monomial symmetric functions of $u,v,w$ (sum of all distinct
permutations of the monomial $u^pv^qw^r$)
\begin{eqnarray}
C_{6} =& m_{(6)} -10 m_{(3,3)}\,,\\
C_9=& (u^3-v^3)(u^3-w^3)(v^3-w^3)\,,\\
C_{12}=& m_{(12)}+4m_{(9,3)}+6m_{(6,6)}+228m_{(6,3,3)}\,.
\end{eqnarray}
It is easily checked by direct calculation that the Jacobian  of these
three functions is nonzero for generic values of $(u,v,w)$. Actually,
its zero set consists of twelve planes, whose geometric significance
will be discussed below.

Let us denote  by $\varphi~: \mathcal{H} \xrightarrow{(I_6,I_9,I_{12})} \C^3$,
the map sending a trilinear form to its three invariants, so that
$(C_6,C_9,C_{12})=\varphi(N_{uvw})$.
Let $S=\{N_{uvw}|(u,v,w)\in \C^3\}$ be the three dimensional space
of normal forms.
The nonvanishing of the Jacobian proves that
$\varphi$ induces a dominant mapping from $S$ to $\C^3$
(that is, the direct image of any non-empty open subset of $S$
contains a non-empty open subset of $\C^3$).
Note that
the independence of $C_6,C_9,C_{12}$ implies the independence of $I_6,I_9,I_{12}$.
Now, Chanler \cite{Chan} has shown that $I_6,I_9,I_{12}$ separate the orbits
in general position.
This proves that the field of rational invariants of $G$ is freely generated
by $I_6, I_9, I_{12}$ (\cite{Encycl}, Lemma 2.1).
As a consequence, $\varphi$ is a \emph{rational quotient}
(\cite{Encycl}, section 2.4) for the action of $G$ on $\mathcal{H}$
(actually, this also implies that $\varphi$ is a categorical quotient,
by \cite{Encycl}, Proposition 2.5 and  Theorem 4.12,
using that $\varphi_{|S}$ is surjective, whence also  $\varphi$).

There exists a non-empty open subset $Y_0$ of $\C^3$ such that the fiber
of  $\varphi$ over each of its points is the closure of an orbit
(\cite{Encycl}, Proposition 2.5).
Let then $U_0=\varphi^{-1}(Y_0)$. This set cuts $S$ since $\varphi_{|S}$ is
dominant.  Let $U_1$ be the union of all orbits having maximal dimension
(a nonempty open set, the function \emph{dimension of the orbit}
being lower semi-continuous).
It is easy to see that $U_1$ intersects $S$
(for instance at $u=1,v=1,w=-1$, whose orbit has dimension $24=\dim G$,
as may be checked by direct calculation).
Let $S_0=U_1 \cap S$, a dense open subset of $S$.
The set $\varphi^{-1} \varphi (S_0)$ thus contains a dense open subset
$U_2$ of $\mathcal{H}$.
One then checks that  $U_0 \cap U_1 \cap U_2$
(a dense open subset, as an intersection of dense open subsets of
an irreducible space) is contained in $GS$.
This proves $\overline{GS}=\mathcal{H}$, that is, the normal
form $N_{uvw}$ is generic.

Let us remark that the above discussion also proves, thanks to
Igusa's theorem (\cite{Encycl}, Theorem 4.12)
that $\C[\mathcal{H}]^G=\C[I_6,I_9,I_{12}]$, that is, the algebra
of invariants is freely generated by Chanler's invariants.

Is is also possible to give a direct proof of
the normal form by using the same technique
as in Ref. \cite{Chan}.  Chanler's method rely on the geometry of
plane cubics,
which will play a prominent role in the sequel.

\section{\label{ellipt}The fundamental cubics}

The trilinear form $f(\xx,\yy,\zz)$ can be encoded in three ways by a $3\times 3$
matrix of linear forms $M_x(\xx), M_y(\yy)$ and $M_z(\zz)$, defined by
\begin{equation}
f(\xx,\yy,\zz)={}^t\yy M_x(\xx)\zz={}^t\xx M_y(\yy)\zz={}^t\xx M_z(\zz)\yy
\end{equation}
and the classification of trilinear forms amounts to the classification
of one of these matrices, say $M_x(\xx)$, up to left and right multiplication
by elements of $SL(3,\C)$ and action of $SL(3,\C)$ on the variable $\xx$.

The most immediate covariants of $f$ are the determinants of
these matrices
\begin{eqnarray}
X(\xx)&=\det M_x(\xx)=\frac16 B_\alpha\,,\\
Y(\yy)&=\det M_y(\yy)=\frac16 B_\beta\,,\\
Z(\zz)&=\det M_z(\zz)=\frac16 B_\gamma\,.
\end{eqnarray}
These are ternary cubic forms, and for generic $f$ the equations
$X(\xx)=0$, etc. will define non singular cubics (elliptic curves) in
$\PP^2$.  It is shown in Ref. \cite{TC} that whenever one of
these curves is elliptic, so are the other two ones,
and moreover, all three are projectively equivalent.
Actually, one can check by direct calculation that they have
the same invariants.
When $f=N_{uvw}$, these three cubics have even the same equation
and are in the  Hesse canonical form \cite{Coo}
\begin{equation}\label{cubics}
X(\xx)=-\phi(x_1^3+x_2^3+x_3^3)+\psi\, x_1x_2x_3= Y(\xx)=Z(\xx)
\end{equation}
where we introduced, following the notation of Ref. \cite{Ma},
\begin{equation}
\phi=uvw\,,\ \ \psi=u^3+v^3+w^3\,.
\end{equation}
The Aronhold invariants of the cubics (\ref{cubics})
are given by
\begin{eqnarray}\label{SandT}
6^4S=&-\phi(\psi^3+(6\phi)^3)\\
6^6T=& (6\phi)^6+20(6\phi^3)\psi^3-8\psi^6\,.
\end{eqnarray}
These are of course invariants of $f$. We recognize that
$6^4S=-C_{12}$, and we introduce an invariant $I_{18}$
such that $C_{18}=I_{18}(N_{uvw})=6^6T$. The three cubics
have the same discriminant $64S^3+T^2$,  known to
be proportional to the hyperdeterminant of $f$ (see Refs. \cite{GKZ,Mi}),
which we normalize as
\begin{equation}
\Delta= 27(64S^3+T^2)\,.
\end{equation}
Then $\Delta={C'}_{12}^3$, where ${C'}_{12}$ is the product of
twelve linear forms
\begin{eqnarray}
{C'}_{12} =& uvw(u+v+w)(\varepsilon u +v+w)(u+\varepsilon v +w)\nonumber\\
&\times (\varepsilon^2 u+\varepsilon v+w)(u+\varepsilon^2 v+w)\nonumber\\
&\times (\varepsilon u+\varepsilon v +w)(\varepsilon^2u+v+w)\nonumber\\
&\times (\varepsilon u+\varepsilon^2 v+w)(\varepsilon^2u+\varepsilon^2v+w)
\end{eqnarray}
where $\varepsilon=e^{2i\pi/3}$,
so that ${C'}_{12}=0$ is the equation in $\PP^2$ of the twelve lines
containing 3 by 3 the nine inflection points of the pencil of cubics
\begin{equation}
u^3+v^3+w^3+6m\, uvw=0\,
\end{equation}
obtained from $X,Y,Z$ by treating the original variables as parameters.
We note also that the Jacobian of $C_6,C_9,C_{12}$ is proportional
to ${C'}_{12}^2$.

\section{\label{groups}Symmetries of the normal forms}

In this section, we will prove Theorem \ref{thnormalform}. That is, a
generic $f$ has 648 different normal forms ( the
cases where this number is reduced will be studied in Section
\ref{formprobl}).

To prove the theorem, we remark that the Hilbert series
(\ref{hilb}) is also the one of the ring of invariants
of $G_{25}$, the group number 25 in the classification of irreducible
complex reflection groups of Shephard and Todd \cite{ShTo}.
This group, which
we will denote for short by $K$, has order 648. It is one of the groups considered
by Maschke \cite{Ma} in his determination of the invariants of the
symmetry group of the 27 lines of a general cubic surface in $\PP^3$
(a group with 51840 elements, which is related to the exceptional
root system $E_6$). To define $K$, we first have to introduce Maschke's
group $H$, a group of order 1296, which is generated by the matrices
of the linear transformations on $\C^3$ given in Table 1.

\begin{table}\label{table1}
\caption{The generators of $H$}
\begin{tabular}{@{}|l||r|r|r|r|l|}
\hline
 & $A$ & $B$ & $C$ & $D$ & $E$ \\
\hline $u'$&$v$& $u$&$u$&$u$ &$\frac{1}{i\sqrt{3}}(u+v+w)$ \\
   $v'$ & $w$ & $w$ & $\varepsilon v$ & $\varepsilon v$ &$\frac{1}{i\sqrt{3}}(u+
 \varepsilon v +\varepsilon^2 w)$\\
$w'$ & $u$ & $v$ & $\varepsilon^2 w$ & $\varepsilon w$&$\frac{1}{i\sqrt{3}}(u+
\varepsilon^2v+\varepsilon w)$\\
\hline
\end{tabular}
\end{table}

This group contains in particular the permutation matrices, and
simultaneous multiplication by $\pm \varepsilon^k$,
since $E^2=-B$. The subgroup
$K$ is the one in which odd permutations can appear only with a
minus sign. It is generated by $A,C,D,E$.

Then, as proved by Maschke, the algebra of invariants of $K$
in $\C[u,v,w]$ is precisely $\C[C_6,C_9,C_{12}]$.

Hence, we can conclude that $K$ is the symmetry group of the normal
forms $N_{uvw}$. There was another, equally natural possibility
leading to the same Hilbert series. The symmetry group $L$
of the equianharmonic cubic surface $\Sigma:\ z_0^3+z_1^3+z_2^3+z_3^3=0$
acting on the homogeneous coordinate ring $\C[\Sigma]$ has as fundamental
invariants the elementary symmetric functions of the $z_i^3$, the first
one being 0 by definition, so that the Hilbert series of $\C[\Sigma]^L$
coincides with (\ref{hilb}). Moreover, $L$ is also of order 648, but
it is known that it is not isomorphic to $K$.

Taking into account the results of Section \ref{genNF}, we see that
\begin{equation}
 S=\{N_{uvw}\,|\, (u,v,w)\in\C^3\}
\end{equation}
is what is usually called
a Chevalley section  of the action of $G$ on ${\cal H}$,
with Weyl group $K$ (see Ref. \cite{Encycl}, p. 174).
This implies that each generic orbits intersects $S$ along a
$K$ orbit, which in turn implies Theorem \ref{thnormalform}.

\section{\label{formprobl}The form problem}

This section contains the proofs of Theorem \ref{coef} and \ref{rsystem}.
We shall formulate 
and solve what
Felix  Klein (see Ref. \cite{Klein})
called the ``Formenproblem''
associated to a finite group action. This is the following: given the numerical
values of the invariants, compute the coordinates of a point
of the corresponding orbit.

In our case,  we shall see that
the problem of finding the parameters $(u,v,w)$
of the normal form of a given generic $f$, given the values
of the invariants, can be reduced to a chain of algebraic equations
of degree at most 4, hence sovable by radicals.

Let $a=I_6$, $b=I_{12}$ and $c=I_{18}$ (we start with $I_{18}$,
because $C_{18}$ is a symmetric function of $u^3,v^3,w^3$, and
at the end of the calculation, select
the solutions which give the correct sign for $C_9$, which is
alternating).

What we have to do is to determine the elementary symmetric functions
$e_1=\psi,e_2=\chi,e_3=\phi^3$ of $u^3,v^3,w^3$. Let $\lambda=216\phi^3$.
Then,
\begin{eqnarray}
\psi^4+\lambda\psi-b &= 0\\
\psi^6-\frac52\lambda\psi^3-\frac18\lambda^2-c &=0\,.
\end{eqnarray}
Eliminating $\lambda$ from these equations, we get a quartic
equation for $\psi^2$
\begin{equation}
27\psi^8-18b\psi^4-8c\psi^2-b^2=0\,.
\end{equation}
The discriminant (with respect to $\psi$) of this polynomial
is proportional to $D=b^2(b^3-c^2)^4$.
When it is non zero, we get 8 values for $e_1$,
each of which
determines univocally $e_2$ and $e_3$.
Hence, we obtain eight cubic equations whose roots are
the possible values of $u^3,v^3,w^3$. This gives 8 sets,
whence $8\times 6=48$ triples, each of which providing
generically 27 values
of $(u,v,w)$, in all $48\times 27=1296$ triples corresponding
to the given values of $a,b,c$, among which exactly $1296/2=648$
give the correct sign for $I_9$.
The common discriminant of the 8 cubics is $\delta=a^3-3ab+2c$.
Clearly, when  $\delta\not=0$, we will have 648 triples.
If $\delta=0$, one can check that the cubics cannot have
a triple root, and that no root is zero. Hence, in this case,
we obtain again 648 triples.

\begin{figure}[t]
\resizebox{7cm}{7cm}{\includegraphics{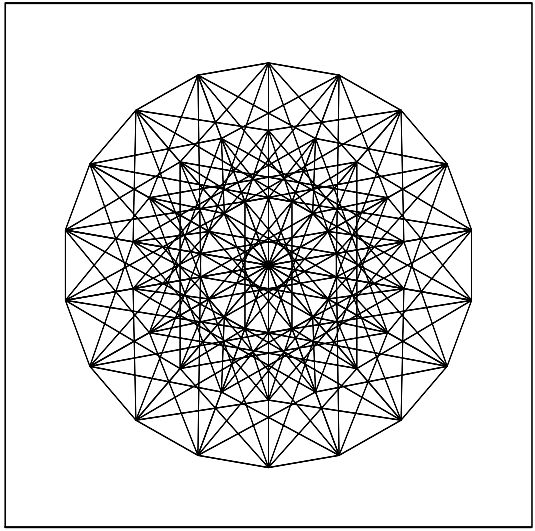}}
\caption{\label{24333}The polyhedron $2\{4\}3\{3\}3$}
\end{figure}

\begin{figure}[t]
\resizebox{7cm}{7cm}{\includegraphics{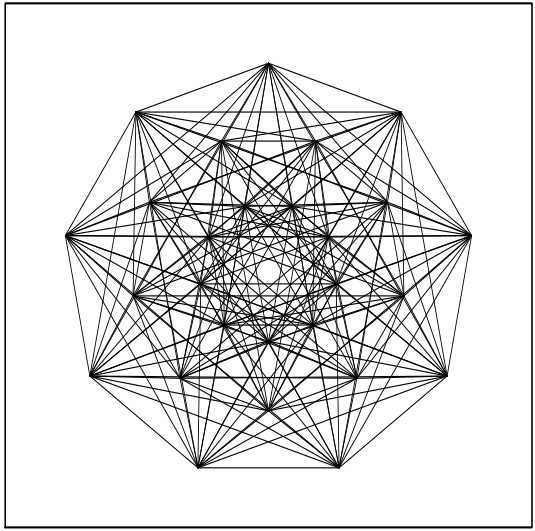}}
\caption{\label{Hessian} The Hessian polyhedron}
\end{figure}

\begin{figure}[t]
\resizebox{7cm}{7cm}{\fbox{\includegraphics{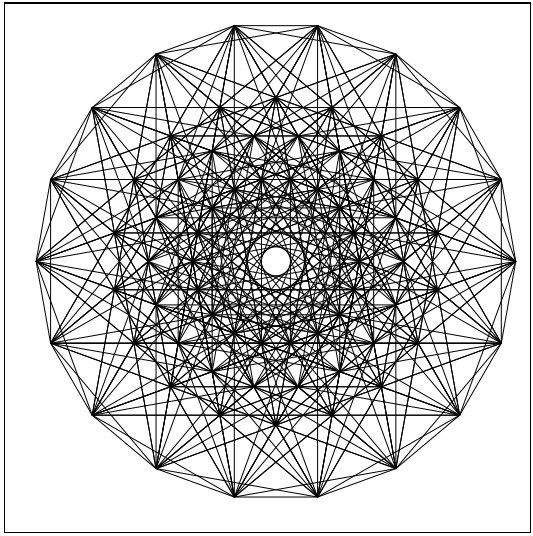}}}
\caption{\label{33342} The  polyhedron $3\{3\}3\{4\}2$}
\end{figure}

If $D=0$, we can have $b^3=c^2$ or $b=0$. In the first
case,
setting $b=q^2,c=q^3$, the equation becomes
\begin{equation}
(\psi^2-q)^3(\psi^2+2q)^3=0\,.
\end{equation}
In this case, we get only four quartics for $\psi^2$. If
$C_9\not=0$, we obtain 216 triples.
If $C_9=0$ and $b=a^2/4$, $c=-a^3/8$ we obtain again $216$ triples which form
the centers of the edges of a complex
polyhedron of type $2\{4\}3\{3\}3$ in $\C^3$
(see Fig. \ref{24333}), in the notation of Ref. \cite{Coxbook}.
The vertices of this polyhedron are the vertices of two reciprocal
Hessian polyhedra (see Fig. \ref{Hessian}) and its edges
 join each vertex of one
Hessian polyhedron to the $8$ closest vertices of the other one.
 In Fig. \ref{Hessian}, the edges of the Hessian polyhedron, which are complex
lines, are represented by real equilateral triangles, so that the
figure can as well be interpreted as a 2-dimensional projection of
a 6-dimensional Gosset polytope $2_{21}$.
If $C_9=0$ and $b=a^2$, $c=a^3$, we
obtain only $72$ triples which are the centers of the edges of a
Hessian polyhedron
and the vertices of a complex polytope of type $3\{3\}3\{4\}2$
(see Fig. \ref{33342}).\\
In the case where $b=0$, we have to
distinguish between the cases $c\not=0$ and $c=0$. If $c\not =0$,
we find 648 triples, whatever the value of $a$. If $c=0$, we
obtain 27 triples if $a\not=0$, and only one if $a=0$.

Indeed, for $b=c=0$, the $\psi$-equation reduces to $\psi^8=0$,
and all the cubics collapse to $12U^3-aU=0$. For $a\not=0$
we obtain precisely 27 triples $(u,v,w)$ which form
the vertices of a Hessian polyhedron in $\C^3$
(see Ref. \cite{Cox}).

From the results of Ref. \cite{OS} about the arrangement
of 12 planes formed by the mirrors of the pseudoreflections
of $K=G_{25}$, we can determine the structure of the stabilizers
of the normal forms. The only nontrivial cases are:
\begin{itemize}
\item the orbits with 216 elements, for which the stabilizer is the
cyclic group $C_3$;
\item the orbits with $72$ elements, for which it
is $C_3\times C_3$;
\item  the Hessian orbits with
27 elements, for which it is the group $G_4$ of the Shephard-Todd
classification.
\end{itemize}

These results can be regarded as a complete description of the moduli
space of three qutrits states. To see what this means, let us recall
some definitions from geometric invariant theory.

It is well known that it in general, the orbits of a group action
on an algebraic variety cannot be regarded as the points of
an algebraic variety. To remedy this situation, one has to discard
certain degenerate orbits. It is then possible to construct
a {\it categorical quotient} and a {\it moduli space},
which describe the geometry of sufficiently generic orbits,
respectively in the affine and projective situation.
          
The {\em categorical quotient} $Y={\cal H}//G$
is defined as the affine variety whose affine coordinate ring
is the ring of polynomial invariants $R=\C[{\cal H}]^G$. The
moduli space is  the projective
variety ${\cal M}={\rm Proj\,}(R)$ of which $R$ is the homogeneous coordinate
ring. It is the quotient of the set
$\PP({\cal H})^{\rm ss}$
of {\em semi-stable} points by the action of $G$ 
(by definition, a point is semi-stable iff at least
one of its algebraic invariants is nonzero, see \cite{Encycl}).

Now, since in our case the algebra of invariants is a polynomial algebra,
we see that the categorical quotient is  just the affine space $\C^3$.
 
The moduli space is more interesting. The projective
variety whose homogeneous coordinate ring is a polynomial
algebra over generators of respective degrees $d_1,\ldots,d_m$
is called a {\em weighted projective space}
$\PP(d_1,\ldots,d_m)$.
Hence, by definition, our moduli space ${\cal M}$ is the weighted projective space
$\PP(6,9,12) \simeq \PP(2,3,4)$.
It is known that this space is isomorphic to
$\PP(1,2,3)$ \cite{Dol},
which in turn can be embedded  as a sextic surface in $\PP^6$,
the so-called {\em del Pezzo surface} $F^6$ (see Ref. \cite{Har}).
The del Pezzo surfaces are very interesting
objects, known to be related to
the exceptional root systems (see, e.g., Ref. \cite{Man}).

The above results can then be interpreted as a description of
the singularities of
${\cal M}$, since one can view it as the quotient of the projective plane
$\PP^2$ of the parameters $(u:v:w)$ under the projective action
of $G_{25}$. We have described this quotient as a 
648-fold ramified covering $\PP^2\rightarrow {\cal M}$, and analyzed
its ramification locus.

\section{\label{concl}Conclusion}

A problem of current interest in Quantum
Information Theory has been connected to various
important mathematical works, scattered on a period
of more than one century
from Ref. \cite{Ma} in 1889 to Ref. \cite{Nur2} in 2000,
in general independent of each other and apparently
discussing different subjects.
Relying on all these works,
we have described the geometry  of the
normal forms of semi-stable orbits of three qutrit
states
under the action of $SL(3,\C)^{\times 3}$, the group of
local filtering
(SLOCC) operations. From a physical point a view, our results
can be expected to
provide a good  starting point for studying the richness of the
entanglement of three qutrits and its differences with that of the
simpler qubit systems. From a mathematical point a view,
we have worked out an interesting example of a problem in
invariant theory, using both classical algebraic and
modern geometric methods,
found a surprising connection with the geometry of complex
polytopes, and applied Klein's vision of Galois theory to
the explicit solution of an algebraic equation of degree 648.

Also, this example provides a good illustration of
the  ideas presented in Refs. \cite{VDD01a} and \cite{Klyachko}.




\bibliographystyle{unsrt}
\bibliography{qutrits5}

\end{document}